\documentclass[reprint,aps,pra]{revtex4-2}

\pdfoutput=1

\usepackage{amsmath,amssymb,hyperref,graphicx,xcolor,marginnote}
\usepackage{epstopdf}
\hypersetup{colorlinks=true,linkcolor=blue,citecolor=blue,urlcolor=blue}
\usepackage{newtxtext,newtxmath}

\renewcommand{\vec}[1]{\mathbf{#1}}
\newcommand{\etarel}{\eta_{\text{rel}}}

\let\originalleft\left
\let\originalright\right
\renewcommand{\left}{\mathopen{}\mathclose\bgroup\originalleft}
\renewcommand{\right}{\aftergroup\egroup\originalright}

\begin{document}
\frenchspacing

%\title{Effect of cavity confinement on the electron-positron contact density in ground-state positronium}
\title{Self annihilation of confined positronium}
\author{A. R. Swann}\email{a.swann@qub.ac.uk}
\affiliation{School of Mathematics and Physics, Queen's University Belfast, University Road, Belfast BT7 1NN, United Kingdom}
\author{D. G. Green}\email{d.green@qub.ac.uk}
\affiliation{School of Mathematics and Physics, Queen's University Belfast, University Road, Belfast BT7 1NN, United Kingdom}
\author{G. F. Gribakin}\email{g.gribakin@qub.ac.uk}
\affiliation{School of Mathematics and Physics, Queen's University Belfast, University Road, Belfast BT7 1NN, United Kingdom}
\date{\today}

\begin{abstract}
The effect of confinement on the self-annihilation rate of positronium  is studied in three levels of approximation. 
Artificial restriction of the electron-positron separation leads to an increase in the annihilation rate over its vacuum value; this  increase is found to diminish exponentially as the maximum separation is increased.
Confinement  in a hard-wall spherical cavity with the center of mass free to move throughout the cavity also increases the annihilation rate over its vacuum value;  the increase depends weakly on the position of the center of mass,  being larger when the center of mass is near the cavity wall.
Finally, to model confinement in a pore of a microporous material, the hard wall is replaced by physically motivated electron- and positron-wall potentials; it is found that the annihilation rate is larger than its vacuum value, in contradiction to calculations of Marlotti Tanzi \textit{et al.} [\href{https://doi.org/10.1103/PhysRevLett.116.033401}{Phys. Rev. Lett. \textbf{116}, 033401 (2016)}] 
that assumed hard-wall confinement for the electrons, and experimental data.
\end{abstract}

\maketitle

\section{\label{sec:intro}Introduction}

Positronium (Ps) is an exotic atom consisting of an electron and a positron.
The expected lifetime of ground-state Ps before annihilation depends on the total spin of its constituent particles; the singlet spin state, parapositronium ($p$-Ps), decays predominantly into two $\gamma$ rays with a lifetime \textit{in vacuo} of $\tau_2\approx0.125$~ns, while the triplet spin state, orthopositronium ($o$-Ps), decays predominantly into three $\gamma$ rays with a lifetime \textit{in vacuo} of $\tau_3\approx142$~ns \cite{Jean88a}.

Confined Ps is used in condensed-matter physics to
estimate pore sizes in microporous materials through positron annihilation
lifetime spectroscopy (PALS); the underlying principle is that confinement of Ps in the pores leads to \textit{pickoff} annihilation of the positron in Ps with an electron in the bulk, 
%if a relative singlet state is achieved, 
the rate of which can be observed and is larger for smaller pores \cite{Gidley06}. 
When Ps exists within a liquid,  exchange repulsion between the Ps electron and the electrons in the surrounding atoms or molecules can lead to the formation of an effective ``bubble'' around the Ps \cite{Ferrel56,Ferrel57}, whose radius can be estimated by PALS analysis.
Confinement of Ps in porous materials also enabled a number of fundamental studies, viz., measurement of Ps-Ps interactions \cite{Cassidy11}, detection of the Ps$_2$ molecule \cite{Cassidy07,Cassidy12}, and measurements of the cavity-induced shift of the Ps Lyman-$\alpha$ transition \cite{Cassidy11a}.
 It is hoped that it may be feasible to use confinement to create a Bose-Einstein condensate of Ps atoms and a $\gamma$-ray laser \cite{Cassidy07a}.
Confinement of Ps in a cavity has also been recently used in combination with many-body theory as a theoretical tool to calculate accurate, \textit{ab initio} elastic scattering cross sections and pickoff annihilation rates in low-energy Ps collisions with many-electron atoms \cite{Swann18,Green18,PsMBTarxiv}.
More generally, the subject of confined atoms is an old one \cite{Michels37,Sommerfield38,deGroot46} that has
seen renewed interest in recent years \cite{Jaskolski96,Buchachenko01,Connerade03,Connerade05,Sabin09,Sabin09a}. Studies
in this area do not serve only as interesting thought experiments but also
elucidate physical situations, e.g., atoms under high pressure \cite{Lawrence81,Connerade00} or trapped in fullerenes \cite{Bethune93,Shinohara00,Komatsu05}. 

The most common model for pore-size
estimation in PALS analysis is the Tao-Eldrup model \cite{Tao72,Eldrup81}, which is valid for pores of radius ${\lesssim}1$~nm. In this model, an $o$-Ps atom in a pore is treated as a point particle of mass $2m_e$ ($m_e$ being the electron mass) moving in a hard-wall  spherical cavity of radius $R_0$.  Because the Ps wave function goes to zero at the wall, there is no overlap with electrons in the bulk, and pickoff annihilation is absent. To mitigate this, it is  postulated that the bulk electrons penetrate a small distance $\Delta R\equiv R_0-R$ into the cavity, where $R$ is the quantity determined to be the radius of the pore.
The pickoff annihilation rate $\lambda_\text{pick}$  is then given  by
\begin{equation}
{\lambda_\text{pick}} =  \lambda_0\left(1-\frac{R}{R+\Delta R} + \frac{1}{2\pi}\sin\frac{2\pi R}{R+\Delta R}\right) ,
\end{equation}
where $\lambda_0=(1/\tau_2+3/\tau_3)/4\approx2.0$~ns$^{-1}$ is the spin-averaged annihilation rate of Ps in vacuum, and
  $\Delta R$ has an empirically determined value of 0.166~nm \cite{Nakanishi88,Nakanishi88a}. 
  Measurement of $\lambda_\text{pick}$ enables $R$ to be determined.
  Note that $\lambda_\text{pick}\to0$ as $R\to\infty$.
  However, it is also possible for the positron  to annihilate with the electron in the $o$-Ps  itself (\textit{self} annihilation), so for larger pore radii, where self annihilation becomes nonnegligible in comparison to pickoff annihilation (${\lesssim}100$~nm), the Tao-Eldrup model can be modified to give the total annihilation rate  as $\lambda=\lambda_\text{pick}+\lambda_\text{self}$, where $\lambda_\text{self}=1/\tau_3$,
%\begin{equation}\label{eq:MTE}
%\lambda=\lambda_\text{pick}+\lambda_\text{self}=\lambda_0 \left( 1-\frac{R}{R+\Delta R} + \frac{1}{2\pi}\sin\frac{2\pi R}{R+\Delta R} \right)+\frac{1}{\tau_3},
%\end{equation}
so that $\lambda\to\lambda_\text{self}$ as $R\to\infty$ \cite{Ito99,MTE_footnote}.
While this is a significant improvement over the original Tao-Eldrup model, it assumes that $\lambda_\text{self}$ has the same value in a pore of any radius as it does in vacuum. In reality, however, the self-annihilation rate is affected by the confinement of the $o$-Ps. The effect of confinement on the self-annihilation rate of $o$-Ps is the subject of this work.

%the self-annihilation rate depends on the overlap of the electron and positron densities within the Ps, i.e., the electron-positron \textit{contact density}, whose value depends on the environment in which the Ps is located.
In the nonrelativistic approximation, the rate of self annihilation  of an $o$-Ps atom into three $\gamma$ rays is given by
\begin{equation}
\lambda_\text{self} = \frac{16}{9}(\pi^2-9)\alpha  r_0^2 c \eta,
\end{equation}
where $\alpha$ is the fine-structure constant, $r_0$ is the classical electron radius,  $c$ is the speed of light, and
$\eta$ is the electron-positron \textit{contact density}, i.e., the density of the positron at the position of the electron \cite{LandauQE}. In vacuum, $\eta=\eta_0\equiv1/(8\pi a_0^3)$, where $a_0$ is the Bohr radius. 
However, in a pore the interactions of the electron and positron with  the bulk may result in $\eta\ne\eta_0$,  the actual value of $\eta$ being determined by two competing effects:
(a) confinement of the $o$-Ps atom to a finite volume of space tends to increase  the contact density;
(b) electric fields arising from  the bulk tend to polarize the $o$-Ps, reducing the contact density   \cite{Dupasquier83,McMullen83}.
If effect (a) is dominant, then the $o$-Ps  is ``compressed'' and $\eta>\eta_0$, while if effect (b) is dominant, then  the $o$-Ps  is ``stretched'' and $\eta<\eta_0$.
In experiment, the $o$-Ps contact density in a particular porous material can be determined by resolving a time annihilation spectrum \cite{Consolati91}. In molecular solids, it is usually found that $\eta<\eta_0$ \cite{Consolati88,Consolati88a,Consolati90,Consolati91a,Consolati93,Nagashima01}, indicating that effect (b) is usually dominant, but values of $\eta>\eta_0$ may also be possible \cite{Consolati91a}. 

Because the Tao-Eldrup model and modifications thereof \cite{Goworek98,Ito99,Gidley99,Dull01,Dutta04,Sudarshan07,Wada13}  treat the Ps  as a point particle, they cannot account for the effects of the confinement on the contact density.
 Consolati \textit{et al.} \cite{Consolati14} have considered Ps as a \textit{bona fide} electron-positron pair in a hard-wall spherical cavity with the distance between the electron and positron artifically restricted to have a maximum value. They demonstrated that the contact density is strongly increased from its vacuum value for maximum separations ${\lesssim}0.3$~nm. 
 %but did not account for interaction of the electron and positron with the wall in a physically meaningful way; they therefore did not attempt to reconcile their calculations with the experimental data. 
Stepanov \textit{et al.} \cite{Stepanov13} considered Ps bubbles in various liquids, with the electron- and positron-liquid interactions described using experimental values for the electron work function, assuming that the positron-liquid work function is equal to the electron-liquid work function. A 10\% reduction in the contact density from its vacuum value was demonstrated for water, which is  smaller than the 25--35\% reduction observed in experiment \cite{Stepanov11}, suggesting that the electron and positron  interact with the liquid differently. Marlotti Tanzi \textit{et al.} \cite{Tanzi16} formulated a model where the electron is strictly confined within the cavity, but the positron is attracted into the bulk (i.e., the positron work function is positive, as is predicted by theoretical models for positrons in condensed matter \cite{Puska94,Bouarissa95,Rubaszek95} and found for silica \cite{Nagashima98}). They also assumed that the cavity had a different effective radius for the electron and positron. 
A reduction in the contact density was observed.

Some assumptions in the models used in the aforementioned works \cite{Consolati14,Stepanov13,Tanzi16}, viz., using a hard-wall cavity, using equal electron and positron work functions, or using different pore radii for the electron and positron, limit the ability of the results  to explain the experimental data that show $\eta<\eta_0$ in most materials \cite{Consolati88,Consolati88a,Consolati90,Consolati91a,Consolati93,Nagashima01}.
In the present work, we investigate the effect of confinement on the contact density in Ps in three levels of approximation: 
(1) a Ps atom is considered in isolation, with the  distance between the electron and positron artificially restricted;
(2) a Ps atom is confined in a hard-wall spherical cavity, with the Ps center of mass able to move freely; 
(3) a Ps atom is confined in a spherical cavity, with the Ps center of mass able to move freely, and interactions of the electron and positron in the Ps with the cavity wall are modeled using Woods-Saxon potentials with distinct electron and positron work functions.
It is our belief that approximation (3) provides a more physical model of Ps confined in a microporous material than previous works \cite{Consolati14,Stepanov13,Tanzi16}.

% \red{we believe that the strict confinement of the electron and the different effective cavity radii are not well justified and that a better model would have the electron and positron moving in a cavity with the same effective radius for each particle, with the electron and positron interacting with the wall through unequal work functions}.\marginnote{Too critical? We just do our own calculation without criticizing others.} In this paper we set up such a model and use it to study the effect of confinement on the Ps contact density for some values of the cavity radius and work functions.

The paper is structured as follows. 
%Section \ref{sec:confined_coulomb} details approximation (1); an analytical formula is obtained for the contact density in an asymptotically large cavity. 
Sections \ref{sec:confined_coulomb}, \ref{sec:hard_wall}, and \ref{sec:soft_wall} detail approximation (1), (2), and (3), respectively.
We conclude in Sec.~\ref{sec:conclusions} with a summary. Except where otherwise stated, we work in atomic units (a.u.) throughout.

\section{\label{sec:confined_coulomb}P\lowercase{s} with restricted electron-positron separation}

Consider a particle of mass $\mu$ moving in an attractive Coulomb field $U(r)=-Z/r$ ($Z>0$). The wave function of a stationary state with orbital quantum number $l$, magnetic quantum number $m$, and energy $E$ is 
\begin{equation}\label{eq:coulomb_wfn}
\psi_{Elm}(\mathbf r)=A_{El} \rho^l e^{-\rho/2}M(l+1-n,2l+2,\rho) Y_{lm}(\vec{\hat{r}}),
\end{equation}
where $A_{El}$ is a normalization constant, $\rho=2\mu Z r/n$, $n=(-2E/\mu Z^2)^{-1/2}$, $M$ is Kummer's confluent hypergeometric function, and $Y_{lm}$ is a spherical harmonic. 
For $E<0$ (i.e., bound states), $n$ is a  positive real number. The  condition $\psi_{Elm}(\mathbf r)\to0$ as $r\to\infty$ requires the quantity $l+1-n$ to take zero or negative-integer values, i.e., the allowed values of $n$ are integers such that $n\ge l+1$. Thus, for a given $l$ and $m$, there are infinitely many bound levels, with energies $E=-\mu Z^2/2n^2$. The wave functions of these levels decay exponentially as $r\to\infty$.
For bound levels with $l=0$, the density of the particle at the origin, is 
\begin{equation}
\lvert\psi_{E00}(\vec0)\rvert^2 =\frac{ \lvert A_{E0}\rvert^2}{4\pi} = \frac{1}{\pi} \left(\frac{\mu Z}{n}  \right)^3,
\end{equation}
while for bound levels with $l>0$, the density of the particle at the origin is zero.  
For $E>0$, the spectrum of energies is continuous and extends from zero to infinity, $n$ and $\rho$ are  imaginary numbers, and the wave functions are oscillatory for large $r$ \cite{LandauQM}.

The states of the relative motion of an electron-positron pair are as described above, where $Z=1$, and $\mu=\frac12$ is the reduced mass of the system. Negative-energy states describe Ps, while positive-energy states describe an unbound electron-positron pair. The energy of the Ps ground state ($l=m=0$, $n=1$) is $-\frac14$~a.u., and the density of this state at the origin, i.e., the electron-positron contact density, is $1/8\pi$~a.u.

Now consider a particle of mass $\mu$ moving in the  potential
\begin{equation}
\widetilde U(r)=
\begin{cases}
-Z/r &\text{if}\quad r<R,\\
\infty &\text{if}\quad r\ge R,
\end{cases}
\end{equation}
where $Z>0$ and $R>0$. The particle is moving in an attractive Coulomb field, with the restriction that it cannot move further than a distance $R$ from the origin. In the region $r<R$, the wave function of a stationary state is given by Eq.~(\ref{eq:coulomb_wfn}), subject to the  boundary condition
\begin{equation}\label{eq:boundary_condition}
M\left(l+1-n,2l+2,\frac{2\mu ZR}{n}\right)=0.
\end{equation}
For a given $l$, solutions of Eq.~(\ref{eq:boundary_condition}) for $n$ (which is no longer constrained to be an integer) can be obtained numerically. 
Restricting our interest to bound states (i.e., positive values of $n$) \footnote{The positive-energy states can be found by solving Eq.~(\ref{eq:boundary_condition}) numerically for imaginary $n$. It will be found that the spectrum of positive-energy levels is infinite but discrete. Although these levels are ``unbound'' in the sense that the particle is not bound in the Coulomb field, the particle is still confined to the region $r<R$, and the wave functions of these levels are  square integrable.}, it will be found that Eq.~(\ref{eq:boundary_condition}) has either no solutions or a finite set of solutions, i.e., the spectrum of bound levels is either empty or finite. The cardinality of the spectrum depends on the value of $R$. For  fixed $l$, the minimum value of $R$ for  $N$ bound levels to exist, which we shall call $R=R_{l,N}$, is found by identifying the $N$th positive root $x$ of the function $M(l+1-n,2l+2,x)$ with $2\mu Z R_{l,N}/n$, where $n\to\infty$ (i.e, $E\to0^-$). In general, for $a\ll-1$ and $b>0$, the $N$th positive root of $M(a,b,x)$ is given by
$
x_N \sim j^2_{b-1,N}/(2b-4a),
$
where $j_{k,N}$ is the $N$th positive root of the Bessel function $J_k$ \cite{AbramowitzStegun}.
Thus, 
$
j^2_{2l+1,N}/4n\sim2\mu ZR_{l,N}/n  
$
for $n\to\infty$, which gives
\begin{equation}\label{eq:Rmin}
R_{l,N}= \frac{j_{2l+1,N}^2}{8\mu Z}.
\end{equation}

By setting $Z=1$ and $\mu=\frac12$, the energy levels can be determined for a Ps atom in which the separation between the electron and positron, $r$, is artificially restricted to values less than $R$.
 From this point onwards, we restrict our interest to the ground state of Ps, whence $l=m=0$. 
For a given value of $R$, the corresponding value of $n$ for the ground state is the first positive solution of Eq.~(\ref{eq:boundary_condition}).
%may be found by identifying $R/n$ with the first positive root $x$ of the function $M(1-n,2,x)$. 
%The ground-state energy is still given by $E=-1/4n^2$.
 %but with $n$ no longer being necessarily integral. } 
In the asymptotic limit $R\to\infty$, we expect  $n\to1$, so if we define an energy correction $\Delta E$ by
\begin{equation}\label{eq:energy_def}
E=-\frac1{4n^2}\equiv-\frac{1}{4} + \Delta E,
\end{equation}
we expect to find $\Delta E\to0$ as $R\to\infty$.
The smallest value of $R$ for which the bound ground state exists is $R_{0,1}=j_{1,1}^2/4\approx3.67$~a.u. [see Eq.~(\ref{eq:Rmin})];
note that in Ref.~\cite{Consolati14} it was incorrectly claimed that $R_{0,1}=\pi\text{~a.u.}$
The electron-positron contact density is given by $\eta\equiv\lvert\psi(\vec0)\rvert^2$ (dropping the subscripts $n$, $l$, and $m$ for brevity).
To compute it, one needs the value of the normalization constant. Since the ground-state wave function is 
\begin{equation}
\psi(\vec{r}) = A e^{-r/2n} M\left(1-n,2,\frac{r}{n}\right) \frac{1}{\sqrt{4\pi}}
\end{equation}
(where $Y_{00}=1/\sqrt{4\pi}$ has been used), we have
\begin{equation}\label{eq:1overA2}
\frac{1}{\lvert A\rvert^2} = \int_0^{R} e^{-r/n} M^2 \left(1-n,2,\frac{r}{n} \right) r^2 \, dr.
\end{equation}
The contact density is then given by
\begin{equation}
\eta = \frac{\lvert A\rvert^2}{4\pi}.
\end{equation}
We can also define the \textit{relative} contact density $\etarel$ as the ratio of  $\eta$ to the corresponding value for $R\to\infty$, viz., $\eta_0\equiv1/8\pi$~a.u.:
\begin{equation}\label{eq:etarel}
\etarel = \frac{\eta}{\eta_0} = 8\pi\eta = 2 \lvert A \rvert^2.
\end{equation}
A value of $\etarel<1$ ($\etarel>1$) indicates that the Ps is stretched (compressed). If we define a relative contact density correction $\Delta\etarel$ by
\begin{equation}\label{eq:Deltaetarel}
\etarel \equiv 1 + \Delta\etarel,
\end{equation}
then we expect to find $\Delta\etarel\to0$ as $R\to\infty$.

While it is straightforward to calculate $\Delta E$ and $\Delta\eta_\text{rel}$ for a particular value of $R$ numerically, %[by solving Eq.~(\ref{eq:boundary_condition}) for $n$ numerically and using Eqs.~(\ref{eq:energy_def}), (\ref{eq:1overA2}), and (\ref{eq:etarel})],
 it is also informative to seek analytical formulae for $\Delta E$ and $\Delta\eta_\text{rel}$ for large $R$ and hence determine the type of decay $\Delta E$ and $\Delta\etarel$ exhibit as $R\to\infty$. The essentially equivalent problem of a  confined hydrogen atom has been under investigation since the 1930s \cite{Michels37,Sommerfield38,deGroot46,Wigner54,Trees56,Dalgarno56,Gray75,Ley-Koo79,Aquino95}, and a rigorous asymptotic formula for the energy correction was obtained by Laughlin \textit{et al.} in 2002 \cite{Laughlin02}. As far as we are aware, an asymptotic formula for the correction to the (relative) contact density has not so far been discovered. In Appendix~\ref{sec:asymp_shifts} we derive such a formula and in the process find the next-order term in the formula for the energy correction in Ref. \cite{Laughlin02}. Here we  simply state the results for convenience:
\begin{align}
\label{eq:DeltaE_expr}
\Delta E &\simeq \frac12 \left( R^2 - 2R - 2 - \frac{8}{R} \right) e^{-R}, \\
\label{eq:con_dens_approx}
\Delta\etarel &\simeq  \big[ R^3 + 2(-4+\gamma+\ln R)R^2 \nonumber\\
&\quad{}+ 4(1-\gamma-\ln R)R - 4(\gamma + \ln R)\big] e^{-R} ,
\end{align}
where $\gamma=0.5772\dots$ is the Euler-Mascheroni constant.
The expression for $\Delta E$ improves on that of Laughlin \textit{et al.} \cite{Laughlin02} by including the $O(R^{-1}e^{-R})$ term explicitly. We note that $\Delta E$ and $\Delta\etarel$ decay exponentially as $R\to\infty$.

\begin{figure}
\centering
\includegraphics{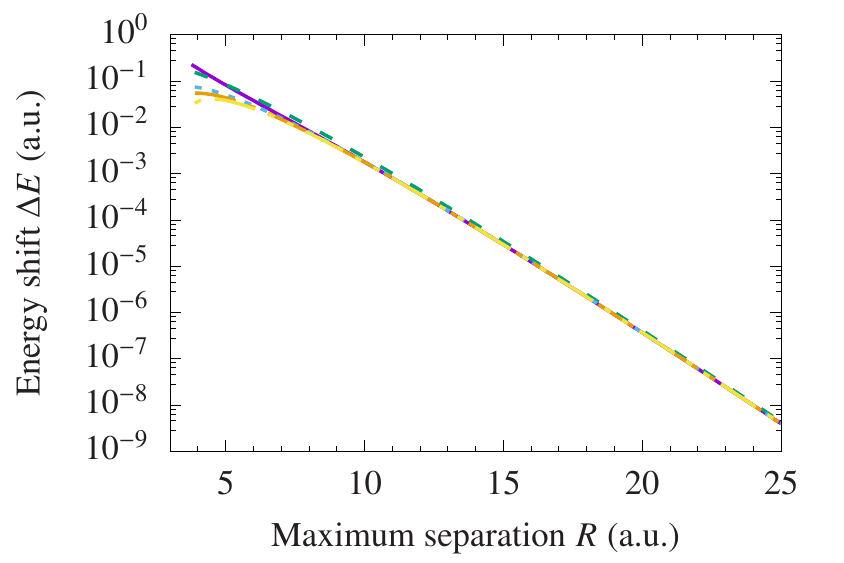}
\caption{\label{fig:DeltaE}Dependence of the energy shift $\Delta E$ on the maximum electron-positron separation $R$. Solid purple line, exact [numerical solution of Eq.~(\ref{eq:boundary_condition})]; short-dashed green line, Eq.~(\ref{eq:DeltaE_expr}) with $R^2 e^{-R}$ term only; dotted blue line, Eq.~(\ref{eq:DeltaE_expr}) with $R^2 e^{-R}$ and $R e^{-R}$ terms only; dash-dotted orange line, Eq.~(\ref{eq:DeltaE_expr}) with $R^2 e^{-R}$, $R e^{-R}$, and $ e^{-R}$ terms only; dash-double-dotted yellow line, Eq.~(\ref{eq:DeltaE_expr}) with all four terms.}
\end{figure}

Figure~\ref{fig:DeltaE} shows the dependence of $\Delta E$ on $R$, where the calculation has been done both numerically, and approximately using Eq.~(\ref{eq:DeltaE_expr}).
For small $R$, Eq.~(\ref{eq:DeltaE_expr}) with only the first term included (i.e., $\Delta E \simeq R^2 e^{-R}/2$) gives the best approximation to the exact result. This is because the asymptotic series (\ref{eq:Ei_asymp}) diverges more quickly for small $R$ than for large $R$, and for small $R$ it is more accurate when fewer terms are included. Conversely, Eq.~(\ref{eq:DeltaE_expr}) with all four terms included gives more accurate results at large $R$. 
%\red{The shift in the energy relative to its vacuum value of $-\frac14$~a.u. is less than 10\% for $R\gtrsim 6.5$~a.u. and less than 1\% for $R\gtrsim 9.5$~a.u.}
The magnitude of the shift in the energy relative to the $R\to\infty$ value of $-\frac14$~a.u. is given by $4\,\Delta E$, and this is less than 10\% for $R\gtrsim7$~a.u. and less than 1\% for $R\gtrsim10$~a.u.

\begin{figure}
\centering
\includegraphics{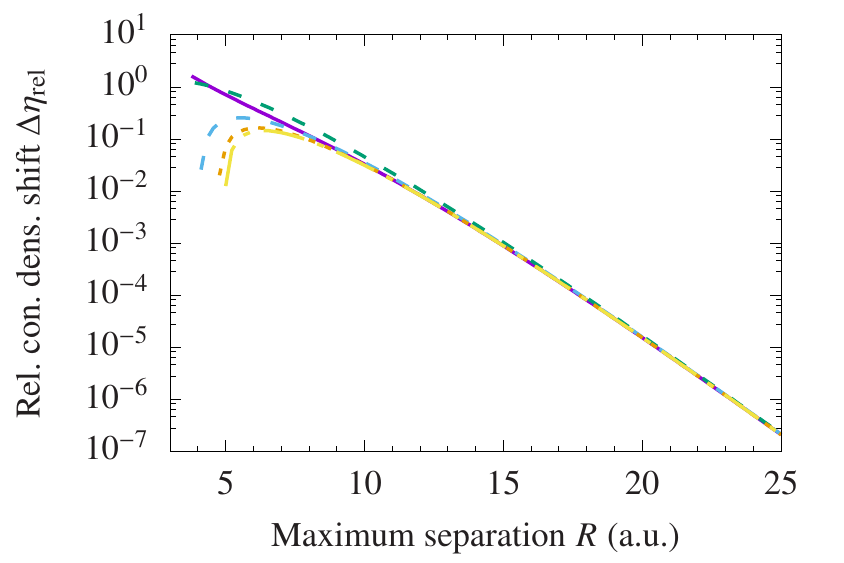}
\caption{\label{fig:Delta_eta_rel}Dependence of the relative contact density shift on the cavity radius. Solid purple line, exact [numerical solution of Eq.~(\ref{eq:1overA2})]; short-dashed green line, Eq.~(\ref{eq:con_dens_approx}) with $R^3 e^{-R}$ term only; dotted blue line, Eq.~(\ref{eq:con_dens_approx}) with $R^3 e^{-R}$ and $R^2 e^{-R}$ terms only; dash-dotted orange line, Eq.~(\ref{eq:con_dens_approx}) with $R^3 e^{-R}$, $R^2 e^{-R}$, and $R e^{-R}$ terms only; dash-double-dotted yellow line, Eq.~(\ref{eq:con_dens_approx}) with all terms.}
\end{figure}

Figure~\ref{fig:Delta_eta_rel} shows the dependence of  $\Delta\eta_\text{rel}$ on $R$. Again, the calculation has been done numerically using Eq.~(\ref{eq:1overA2}), and approximately using Eq.~(\ref{eq:con_dens_approx}).
For small $R$, the best approximation is again given by including only the first term in Eq.~(\ref{eq:con_dens_approx}), while for large $R$, including all terms gives the best result. We note than $\Delta\etarel < 10$\% for $R\gtrsim 8$~a.u., and $\Delta\etarel < 1$\% for $R\gtrsim 12$~a.u.

\section{\label{sec:hard_wall}Hard-wall confinement of P\lowercase{s} with free center-of-mass motion}

\subsection{\label{sec:hard_wall_theory}Theory}

The  Hamiltonian for an electron-positron pair confined in a cavity with an impenetrable spherical wall of radius $R$ is
\begin{equation}
H = h_e(\vec r_e) + h_p(\vec r_p) - V(\vec r_e,\vec r_p) ,
\end{equation}
where $\vec{r}_e$ ($\vec{r}_p$) is the position of the electron (positron) relative to the center of the cavity,
\begin{equation}
h_{e,p}(\vec r) = -\frac12 \nabla_{\vec r}^2 + V_{e,p}(\vec r)
\end{equation}
are the single-particle Hamiltonians for the electron and positron,
 $V_e$ ($V_p$) is the confining potential for the electron (positron),
\begin{equation}\label{eq:conf_pot_inf}
V_{e,p}(\vec r)= 
\begin{cases}
0 & \text{if}\quad r<R , \\
\infty & \text{if}\quad r\ge R,
\end{cases}
\end{equation}
and $V(\mathbf{r}_e,\mathbf{r}_p)=1/\lvert \vec{r}_e - \vec{r}_p\rvert$, so that $-V$ is the attractive Coulomb interaction between the two particles.

Because of the confining potentials (\ref{eq:conf_pot_inf}), the spectra of the single-particle Hamiltonians $h_{e,p}$ are fully discrete.
The eigenstates of $h_{e,p}(\vec r)$  are simply those for a particle moving freely in the region $r<R$. The condition that the wave functions must vanish at $r=R$ gives the energy and wave function for eigenstate $ \mu $ as
\begin{align}
\varepsilon_{\mu} &= \frac{k_\mu^2}{2}= \frac{Z_{l_\mu+1/2,n_\mu}^2}{2R^2} , \\
\varphi_{\mu}(\vec{r}) &=  \frac{1}{r} P_\mu(r) Y_{l_\mu m_\mu}(\vec{\hat{r}}) \quad (r<R),
\end{align}
respectively, where 
\begin{equation}\label{eq:sin_par_rad_wfn}
P_\mu(r) = A_\mu \sqrt{r} J_{l_\mu + 1/2}(k_\mu r) ,
\end{equation}
$J_\nu$ is the Bessel function, $Z_{\nu,j}$ is the $j$th root of $J_\nu$, $k_\mu=Z_{l_\mu+1/2,n_\mu}/R$ is the momentum, and $A_\mu$ is a normalization constant. The radial, orbital, and magnetic quantum numbers are denoted by $n_\mu$, $l_\mu$, and $m_\mu$ respectively. Being eigenstates of a single-particle Hamiltonian, the set of wave functions $\{ \varphi_\mu(\vec{r})\}$ is a complete basis in which a general function of $\vec{r}$ may be expanded.
A two-particle Ps wave function with fixed total angular momentum $J$ and parity $\Pi$ may be expanded in these single-particle wave functions as
\begin{subequations}
\begin{align}\label{eq:Ps_wfn_expansion}
\Psi_{J\Pi} (\vec{r}_e,\vec{r}_p) &= \sum_{\mu\nu} C_{\mu\nu} \varphi_\mu(\vec{r}_e) \varphi_\nu(\vec{r}_p) \\
&\equiv \sum_{n_\nu l_\mu n_\nu l_\nu} C_{n_\mu l_\mu n_\nu l_\nu} \frac{1}{r_e} P_{\mu}(r_e) \frac{1}{r_p} P_{\nu}(r_p) \nonumber\\
&\quad{}\times \sum_{m_\mu m_\nu} C_{l_\mu m_\mu l_\nu m_\nu}^{JM} Y_{l_\mu m_\mu}(\vec{\hat{r}}_e) Y_{l_\nu m_\nu}(\vec{\hat{r}}_p) ,\label{eq:Ps_wfn_expansion2}
\end{align}
\end{subequations}
where the $C_{n_\mu l_\mu n_\nu l_\nu}$ are expansion coefficients and the $C_{l_\mu m_\mu l_\nu m_\nu}^{JM}$ are Clebsch-Gordan coefficients. Substitution of Eq.~(\ref{eq:Ps_wfn_expansion}) into the  Schr\"odinger equation,
%\begin{equation}
$H\Psi_{J\Pi}=E\Psi_{J\Pi}$,
%\end{equation}
leads to an eigenvalue equation,
%\begin{equation}
$\vec{HC}=E\vec{C}$,
%\end{equation}
where the Hamiltonian matrix $\vec{H}$ has elements
\begin{align}
\langle \nu' \mu' \vert H \vert \mu \nu \rangle = \left( \varepsilon_\mu+\varepsilon_\nu \right) \delta_{\mu\mu'} \delta_{\nu\nu'} - \langle \nu' \mu' \vert V \vert \mu \nu \rangle,
\end{align}
and the Coulomb matrix element is defined as
\begin{align}\label{eq:cou_mat_ele}
\langle \nu' \mu' \vert V \vert \mu \nu \rangle &= \iint \left[ \varphi_{\nu'}(\vec{r}_p) \varphi_{\mu'}(\vec{r}_e) \right]^* \nonumber\\
&\quad{}\times \frac{1}{\lvert \vec{r}_e - \vec{r}_p \rvert} \varphi_\mu(\vec{r}_e) \varphi_\nu(\vec{r}_p) \, d^3\vec{r}_e \, d^3\vec{r}_p.
\end{align}
After separating the radial and angular parts of the wave functions in Eq.~(\ref{eq:cou_mat_ele}), integrating over the angular variables analytically, and summing over the magnetic quantum numbers (see Appendix~\ref{sec:integrals}), the Hamiltonian matrix can be diagonalized to obtain the energy eigenvalues $E$ and the expansion coefficients $C_{n_\mu l_\mu n_\nu l_\nu}$.

With Ps states constructed in this way, the Ps center of mass is able to move freely within the cavity. Thus, the electron-positron contact density may depend on the position of the center of mass. The mean value  is given by
\begin{equation}\label{eq:mean_contact_density}
\langle\eta\rangle = \iint \left\lvert \Psi_{J\Pi}(\vec{r}_e,\vec{r}_p) \right\rvert^2 \delta(\vec{r}_e-\vec{r}_p) \, d^3\vec{r}_e \, d^3\vec{r}_p.
\end{equation}
%The angular integration is performed analytically, and the magnetic quantum numbers and spins are summed over. 
One can also calculate the density of the Ps center of mass at an arbitrary position $\vec{r}$ as
\begin{equation}\label{eq:cm_dens_int}
\rho_\text{cm}(\vec{r}) = \iint \left\lvert \Psi_{J\Pi}(\vec{r}_e,\vec{r}_p) \right\rvert^2 \delta\left( \frac{\vec{r}_e+\vec{r}_p}{2} - \vec{r}\right) \, d^3\vec{r}_e \, d^3\vec{r}_p.
\end{equation}
See Appendix~\ref{sec:integrals} for details on how these integrals are computed.
Finally, the total density at position $\vec{r}$ is
\begin{equation}
\rho_\text{tot}(\vec{r}) = \left\lvert \Psi_{J\Pi}(\vec{r},\vec{r}) \right\rvert^2.
\end{equation}
The center-of-mass density and total density are normalized as
\begin{align}
\int \rho_\text{cm}(\vec{r}) \, d^3\vec{r} &= 1, \\
\int \rho_\text{tot}(\vec{r}) \, d^3\vec{r} &= \langle\eta\rangle.\label{eq:norm_rhotot}
\end{align}

Assuming that the internal motion and the center-of-mass motion of the Ps are decoupled, % total wave function could be factorized as an internal-motion wave function multiplied by a center-of-mass-motion wave function,\
 it is instructive to compare the value of $\rho_\text{tot}(\vec{r})$ with that of $\langle \eta \rangle \rho_\text{cm}(\vec{r})$ for a given $\vec{r}$. If $\rho_\text{tot}(\vec{r}) < \langle \eta \rangle \rho_\text{cm}(\vec{r})$, then the  contact density is lower than its average value when the center of mass is at position $\vec{r}$. Conversely, if $\rho_\text{tot}(\vec{r}) > \langle \eta \rangle \rho_\text{cm}(\vec{r})$, then the  contact density is higher than its average value when the center of mass is at position $\vec{r}$.

\subsection{\label{sec:hard_wall_numimp}Numerical implementation}
Although the single-particle radial wave functions $P_\mu$ have  analytical expressions, Eq.~(\ref{eq:sin_par_rad_wfn}), for the purposes of constructing the two-particle Ps wave function and calculating the Coulomb matrix elements it is more convenient to work numerically \footnote{Another reason for working numerically is that when we later introduce physical electron- and positron-wall potentials (see Sec.~\ref{sec:soft_wall}), the radial wave functions $P_\mu$ no longer have a simple analytical form}. We expand the single-particle radial functions in a set of 60 $B$-spline basis functions of order 9  with a linear knot sequence \cite{deBoor01,Bachau01,Brown17}.
%\begin{equation}
%P_\mu(r) = \sum_{j} c_j^{(\mu)} B_j(r).
%\end{equation}
%A set of $60$ $B$ splines of order 9 has been used thoughout. Upon substituting this expansion into the single-particle radial Schr\"odinger equation 
%\begin{equation}
%h^{(\mu)} P_\mu(r)=\varepsilon_\mu P_\mu(r), 
%\end{equation}
%with 
%\begin{equation}\label{eqn:sin_part_hamil}
%h^{(\mu)}=-\frac12 \frac{d^2}{dr^2} + \frac{l_\mu(l_\mu+1)}{2r^2},
%\end{equation} 
%%and projecting with $B_i(r)$, 
%we obtain
%\begin{equation}
%\sum_j h_{ij}^{(\mu)} c_j^{(\mu)} = \varepsilon_\mu \sum_j Q_{ij} c_j^{(\mu)},
%\label{eq:bsp_mat}
%\end{equation}
%where
%\begin{subequations}
%\begin{align}
%h_{ij}^{(\mu)} &= \int_0^R B_i(r) h^{(\mu)} B_j(r) \, dr ,\\
%Q_{ij} &= \int_0^R B_i(r) B_j(r) \, dr.
%\end{align}
%\end{subequations}
%Equation~(\ref{eq:bsp_mat}) is a generalized matrix eigenvalue equation that is solved to obtain the energy eigenvalues $\varepsilon_\mu$ and expansion coefficients $c_j^{(\mu)}$.
A general feature of the use of $B$-spline basis sets is that all of the spline functions, and consequently any function expanded in them, are clamped to zero at some chosen radius---the \textit{box radius} $R_B$. %Since we consider the cavity to have an impenetrable wall, we simply put $R_B=R$.
Simply setting $R_B=R$ conveniently provides the required hard-wall confinement of the electron and positron within the cavity.

All of our calculations have been carried out for $J=0$, $\Pi=+1$. After the Hamiltonian matrix is diagonalized, we calculate the mean contact density, center-of-mass density, and total density only for the lowest-energy state. Then the internal motion and center-of-mass motion will have zero angular momentum, and the respective wave functions will have no radial nodes. Also, $\rho_\text{cm}(\vec{r})$ and $\rho_\text{tot}(\vec{r})$  depend only on the distance $r$ from the center of the cavity and not on the polar or azimuthal angles, for $J=0$.

The sums in Eq.~(\ref{eq:Ps_wfn_expansion2}) theoretically run over infinitely many values of the orbital and radial quantum numbers. In practice we must set upper limits of $l_\text{max}$ and $n_\text{max}$, respectively. This truncation may have a significant effect on the various quantities that we calculate, so we carry out our calculations for a range of values of  $l_\text{max}$ and $n_\text{max}$, viz., $l_\text{max}=16$--20 and $n_\text{max}=15$--20, and extrapolate the results to the limits $l_\text{max}\to\infty$ and $n_\text{max}\to\infty$ \cite{Brown17}. 
%We have performed calculations for $l_\text{max}=16$--20 and $n_\text{max}=15$--20. 
After some investigation, we determined that $\langle\eta\rangle$ and $\rho_\text{tot}(\vec{r})$ converge  as $1/(l_\text{max}+\frac12)$ and $1/n_\text{max}$, and $\rho_\text{cm}(\vec{r})$ converges as $1/(l_\text{max}+\frac12)^3$ and $1/n_\text{max}^3$.
In practice, we found that extrapolating $\rho_\text{cm}(\vec{r})$ produces a negligible change from the calculation using the largest values of $l_\text{max}$ and $n_\text{max}$. 
However, extrapolation is important for $\langle\eta\rangle$ and $\rho_\text{tot}(\vec{r})$; it is carried out by performing least-squares fits of the following bilinear functions to the data:
\begin{align}
\langle\eta\rangle[l_\text{max}, n_\text{max}] &= \langle\eta\rangle[\infty,\infty] + \frac{\alpha}{l_\text{max}+\frac12} + \frac{\beta}{n_\text{max}} \nonumber\\
&\quad{}+ \frac{\gamma}{(l_\text{max}+\frac12)n_\text{max}} , \label{eq:mean_cd_extrp}\\
\rho_\text{tot}(\vec{r})[l_\text{max}, n_\text{max}] &= \rho_\text{tot}(\vec{r})[\infty, \infty] + \frac{\zeta}{l_\text{max}+\frac12} + \frac{\kappa}{n_\text{max}} \nonumber\\
&\quad{}+ \frac{\lambda}{(l_\text{max}+\frac12)n_\text{max}} . \label{eq:rho_tot_extrp}
\end{align}
We note that after extrapolating $\rho_\text{tot}(\vec{r})$ via Eq.~(\ref{eq:rho_tot_extrp}) and plotting as a function of $r$, a ``bump'' is often observed across a small range of values of $r$ (see, e.g., Fig.~\ref{fig:total_density_hard_wall}). This bump is not present in any of the unextrapolated curves for $\rho_\text{tot}(\vec{r})$ and is due to instability of the extrapolation. An alternative extrapolation function that does not produce such a bump is
\begin{equation}\label{eq:rho_tot_extrp_alt}
\rho_\text{tot}(\vec{r})[l_\text{max}, n_\text{max}] = \rho_\text{tot}(\vec{r})[\infty, \infty] + \frac{\zeta}{l_\text{max}+\frac12} + \frac{\kappa}{n_\text{max}} .
\end{equation}
The mean contact density $\langle\eta\rangle$ could also be extrapolated in a manner akin to Eq.~(\ref{eq:rho_tot_extrp_alt}); however, we found that this leads to a reduction of approximately 0.002 a.u. in the value of $\langle\eta\rangle$ for all $R$ in the range we have considered ($4\leq R\leq 16$~a.u.). This change means we obtain $\langle\eta\rangle<\eta_0$~a.u. for $R\gtrsim 12$~a.u., which is not physical. Therefore, we believe that extrapolation via Eq.~(\ref{eq:mean_cd_extrp}) is more accurate.
Regarding $\rho_\text{tot}(\vec{r})$, while Eq.~(\ref{eq:rho_tot_extrp_alt}) apparently provides more robust extrapolation than Eq.~(\ref{eq:rho_tot_extrp}), extrapolation via Eq.~(\ref{eq:rho_tot_extrp}) is probably more accurate than Eq.~(\ref{eq:rho_tot_extrp_alt}) away from the bump.

\subsection{Results}

\begin{figure}
\centering
\includegraphics{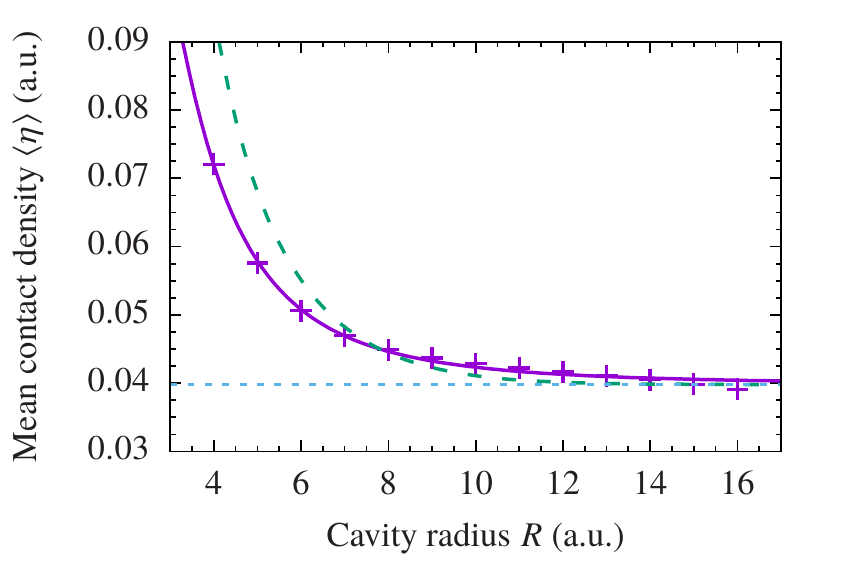}
\caption{\label{fig:contact_hard}Mean contact density of Ps in a hard-wall cavity. Purple plusses, extrapolated via Eq.~(\ref{eq:mean_cd_extrp}); solid purple line, Eq.~(\ref{eq:confit}); dashed green line, numerical result for frozen center of mass (see Sec.~\ref{sec:confined_coulomb}); dotted blue line, vacuum value of $\eta_0=1/8\pi$~a.u.}
\end{figure}

We first calculated $\langle\eta\rangle$ for a range of values of the cavity radius, namely $R=4$--16~a.u. The results are shown in Fig.~\ref{fig:contact_hard} (purple plusses).
As expected, for small $R$ the mean contact density is significantly larger than its vacuum value, but it rapidly decreases as we increase $R$. For $R\gtrsim 14$~a.u., $\langle\eta\rangle$ actually appears to be \textit{lower} than its vacuum value. This is not a physical phenomenon and is probably caused by poor convergence of the Ps wave function (\ref{eq:Ps_wfn_expansion2}) resulting from the large cavity radius. A curve  has been fit to the data points and is also shown in Fig.~\ref{fig:contact_hard} (solid purple line); its equation is
\begin{equation}\label{eq:confit}
\langle\eta\rangle = \eta_0 + \frac{A}{R^3} - \frac{B}{R^5},
\end{equation}
where $A=2.63$ and $B=9.17$~a.u.
See Appendix~\ref{sec:hard_fit} for a simple analysis justifying the analytical form of this fit.
Figure~\ref{fig:contact_hard} also shows the numerical result from Sec.~\ref{sec:confined_coulomb} (dashed green line, same as solid purple line in Fig.~\ref{fig:Delta_eta_rel}), where $R$ was the maximum separation between the electron and positron.
Although the maximum possible separation in the present model is the diameter of the cavity, i.e., $2R$, it is reasonable to make this direct comparison between the two models because in the present model the center of mass is most likely to be found at a distance ${\sim}R/2$ from the center of the cavity (see below), and there the effective maximum separation of the electron and positron is ${\sim}R$.
 %We see that for small $R$, allowing the center of mass to move freely reduces the mean contact density, i.e.,  allows  the contact density to come closer to the vacuum value; this is likely  because allowing the center of mass to move freely lets the Ps ``relax'' into a lower-energy state that is more similar to Ps in a vacuum.
The two models predict a qualitatively similar dependence of the (mean) contact density on $R$.
For large $R$, the contact density tends to $\eta_0$ in both models.
 %but increases it for large $R$; it gives the same result for $R \approx 7.5$~a.u.

\begin{figure}
\centering
\includegraphics{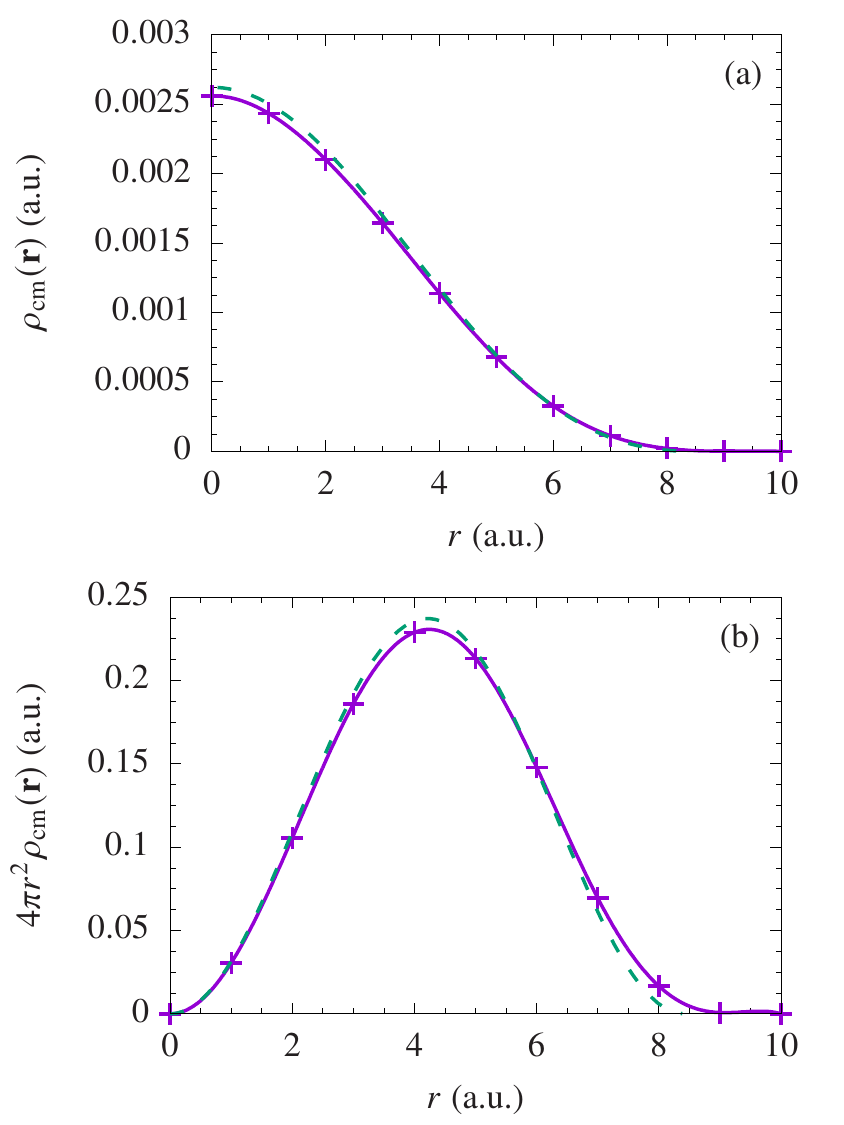}
\caption{\label{fig:COM_density_hard_wall}Center-of-mass density in a hard-wall cavity of radius $R=10$~a.u. Panel (a) shows $\rho_\text{cm}(\vec{r})$ itself, while panel (b) shows $4\pi r^2\rho_\text{cm}(\vec{r})$. Purple plusses, calculations for $l_\text{max}=n_\text{max}=20$; solid purple lines, interpolated from calculations using cubic splines; dashed green lines, Eq.~(\ref{eq:rhocm_sine_wave}) with $K=0.372694$~a.u.}
\end{figure}

To investigate the dependence of the contact density on the position of the center of mass within the cavity, we now calculate $\rho_\text{cm}(\vec r)$ and $\rho_\text{tot}(\vec r)$ for $R=10$~a.u., which has $\langle\eta\rangle=0.0428426$~a.u. Figure~\ref{fig:COM_density_hard_wall} shows $\rho_\text{cm}(\vec{r})$ as a function of $r$ for $l_\text{max}=n_\text{max}=20$ (purple plusses).
This density was calculated for integer values of $r$ in the range $r=0$--10~a.u. and interpolated using cubic splines (solid purple lines).  We checked the normalization of the density by calculating $4\pi \int_0^R  \rho_\text{cm}(\vec{r})r^2dr$ numerically;  we obtained a satisfactory value of 0.999872. We  see from Fig.~\ref{fig:COM_density_hard_wall} that $\rho_\text{cm}(\vec{r})$ actually goes to ${\sim}0$ at some value of $r<R$. This is because the Ps itself has a finite radius, and so the center of mass cannot move all the way to the wall: the center of mass moves in an effectively smaller cavity of radius $R_\text{eff}<R$, where the boundary condition gives $KR_\text{eff}=\pi$, with $K$ the center-of-mass momentum \cite{Brown17}. 

It is useful to compare $\rho_\text{cm}(\vec{r})$ with its expected analytical form. 
Assuming that the internal and center-of-mass motion of the Ps are decoupled,
the center-of-mass wave function is given by the $s$-wave contribution to a plane wave, i.e., $\psi_\text{cm}(\vec{r})=A(Kr)^{-1}\sin (Kr)Y_{00}(\vec{\hat{r}})$, where $A = K\sqrt{2/R_\text{eff}}$ is the normalization constant. Thus,
\begin{equation}\label{eq:rhocm_sine_wave}
\rho_\text{cm}(\vec{r}) = \lvert \psi_\text{cm}(\vec r)\rvert^2 = A^2 \frac{\sin^2 Kr}{(Kr)^2} \frac{1}{4\pi} \qquad (0 < r < R_\text{eff}).
\end{equation}
In our calculation for $l_\text{max}=n_\text{max}=20$, the the energy eigenvalue obtained from diagonalization of the Hamiltonian matrix was $-0.215275$~a.u.; hence $K=0.372694$~a.u. \footnote{The center-of-mass momentum $K$ is estimated from the energy eigenvalue $E$ by assuming that $E=-\frac14+K^2/4$, where $-\frac14$ is the internal energy of ground-state Ps, and $K^2/4$ is the center-of-mass energy.}. Figure~\ref{fig:COM_density_hard_wall} also shows Eq.~(\ref{eq:rhocm_sine_wave}) for this value of $K$ (dashed green lines). There is  very close agreement with the numerical calculation.
Note that the most probable distance of the center of mass from the center of the cavity is  the value of $r$ where $4\pi r^2\rho_\text{cm}(\vec r)$ takes its maximum value. Using Eq.~(\ref{eq:rhocm_sine_wave}), this is $r=R_\text{eff}/2\sim R/2$ (assuming $R_\text{eff}/R\sim1$).

\begin{figure}
\centering
\includegraphics{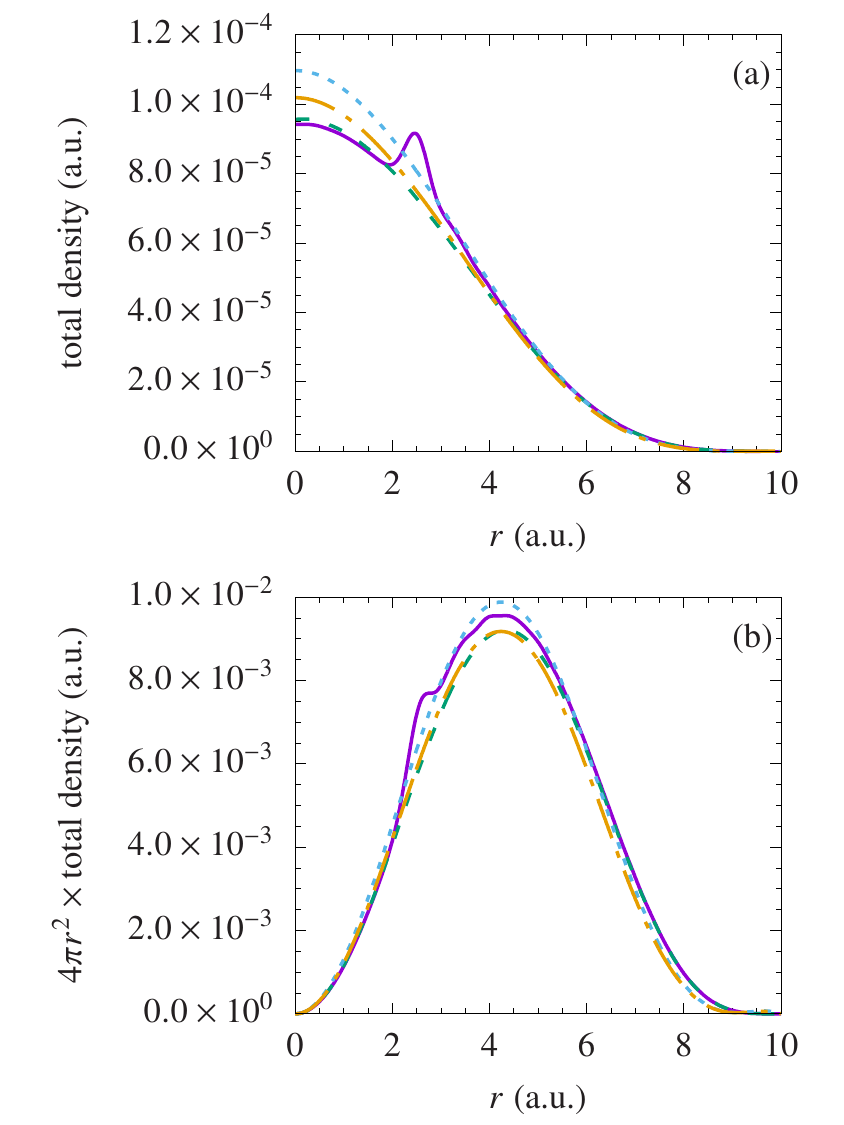}
\caption{\label{fig:total_density_hard_wall}Total density in a hard-wall cavity of radius $R=10$~a.u. Panel (a) shows the total density itself, while panel (b) shows $4\pi r^2$ multiplied by the total density. Solid purple curve, direct calculation of $\rho_\text{tot}(\vec{r})$ with extrapolation via Eq.~(\ref{eq:rho_tot_extrp}); dashed green curve, direct calculation of $\rho_\text{tot}(\vec{r})$ with extrapolation via Eq.~(\ref{eq:rho_tot_extrp_alt}); dotted blue curve, $\langle\eta\rangle \rho_\text{cm}(\vec{r})$; dash-dotted orange curve, $\eta_0\rho_\text{cm}(\vec{r})$.}
\end{figure}

The total density $\rho_\text{tot}(\vec{r})$ is shown in Fig.~\ref{fig:total_density_hard_wall}.
We display the results of extrapolating via Eqs.~(\ref{eq:rho_tot_extrp}) and (\ref{eq:rho_tot_extrp_alt}) separately. It can be seen that extrapolating via Eq.~(\ref{eq:rho_tot_extrp}) produces an unphysical bump at $r\approx2$--3~a.u., while extrapolating via Eq.~(\ref{eq:rho_tot_extrp_alt}) does not. Comparing $\rho_\text{tot}(\vec{r})$ with $\langle\eta\rangle \rho_\text{cm}(\vec{r})$ (also shown), we deduce that $\eta<\langle\eta\rangle$ for $r \lesssim 5$~a.u., $\eta\approx\langle\eta\rangle$ for $r\approx5$--6~a.u., and $\eta>\langle\eta\rangle$ for $r\gtrsim 6$~a.u. This is the expected result: the contact density increases when the Ps approaches (and collides) with the wall. When far from the wall, the Ps is essentially free with a contact density close to the vacuum value. The mean value $\langle\eta\rangle$ results from a tradeoff between these two situations. Figure~\ref{fig:total_density_hard_wall} also shows $\eta_0\rho_\text{cm}(\vec{r})$. We notice that  $\rho_\text{tot}(\vec{r}) < \eta_0\rho_\text{cm}(\vec{r})$ for $r\lesssim 2$~a.u. if using Eq.~(\ref{eq:rho_tot_extrp}) or $r \lesssim 4$~a.u. if using Eq.~(\ref{eq:rho_tot_extrp_alt}). This indicates that the contact density is actually \textit{lower} than its vacuum value near the center of the cavity, i.e., the Ps is stretched. We are unaware of any physical reason why this should occur, and we believe that it is  due to error in the extrapolation of $\rho_\text{tot}(\vec{r})$. Still, at $r=0$, where the effect is greatest, the relative difference between $\rho_\text{tot}(\vec{r})$ and $\eta_0\rho_\text{cm}(\vec{r})$ is only ${\approx}8$\%.

\section{\label{sec:soft_wall}Confinement of P\lowercase{s} with physical electron- and positron-wall interactions}

We now  model confinement of Ps inside a physical pore of a mesoporous material. 
We use the same $B$-spline implementation described in Sec.~\ref{sec:hard_wall}, but rather than having a hard-wall cavity,
we model the interactions of the electron and positron with the wall using Woods-Saxon potentials, viz.,
\begin{equation}
V_{e,p}(\vec{r}) = -\frac{\phi_{e,p}}{1+\exp[(R-r)/\Delta_{e,p}]} , 
\end{equation}
where $\phi_e$ ($\phi_p$) is the electron-wall (positron-wall) work function \footnote{We define the work function to be the amount of energy required to remove the electron or positron from the bulk, so that a positive (negative) work function implies attraction (repulsion) of the electron or positron to (from) the bulk.}, and $\Delta_e$ ($\Delta_p$) is a parameter. Assuming $R/\Delta_{e,p}\gg1$, for $r\ll R$ we have $V_{e,p}(r)\sim0$, for $r=R$ we have $V_{e,p}(r)=-\phi_{e,p}/2$, and for $r\gg R$ we have $V_{e,p}\sim-\phi_{e,p}$. The parameters $\Delta_{e,p}$ characterize the ``width'' of the step in the potentials from 0 to $-\phi_{e,p}$.
To enable a  comparison with the results of Marlotti Tanzi \textit{et al.} \cite{Tanzi16}, we calculate $\langle\eta\rangle$, $\rho_\text{cm}(\vec r)$, and $\rho_\text{tot}(\vec r)$ for
$\phi_p=0.1$~a.u. (${\sim}3$~eV) and $R=10$~a.u. (${\sim}0.5$~nm).
Recall that in the model of Marlotti Tanzi \textit{et al.} \cite{Tanzi16}, the electron was strictly confined within the cavity. This would be equivalent to setting $\phi_e=-\infty$ in the present model.
We instead choose  $\phi_e=-0.5$~a.u. (${\sim}{-}10$~eV) and $\Delta_e=\Delta_p=1$~a.u.
The chosen values of $\phi_{e,p}$ are close to reported experimental values for silica \cite{Griscom77,Trukhin92,Nagashima98}.

Since the electron and positron can now penetrate into the cavity wall, the $B$-spline box radius must be chosen larger than the cavity radius.
The precise value of $R_B$ should not affect the results significantly, provided it is large enough that the potentials $V_{e,p}$  have almost attained their asymptotic values before reaching the box edge. Specifically, for $V_{e,p}$  to be within $q$\% of their asymptotic values at the box edge, we require
\begin{equation}
R_B \geq R + \max(\Delta_e,\Delta_p) \ln \frac{q}{100-q},
\end{equation}
e.g., for $R=10$~a.u., $\Delta_e=\Delta_p=1$~a.u., and $q=95$, we require $R_B \geq 12.9$~a.u. On the other hand, making $R_B$ too large could negatively affect the convergence of the Ps wave function (\ref{eq:Ps_wfn_expansion2}) with respect to the number of partial waves and radial states included, thus making the extrapolation more uncertain. A balance  must  be sought,
so we have chosen to use a box radius of $R_B=15$~a.u.
To verify that the results do not depend significantly on the choice of $R_B$, the calculations were also carried out for $R_B=14$~a.u., and indeed it was found that the change in the results is negligible. 
%Since $\phi_e<0$ and $\phi_p>0$, the electron will tend to remain confined in the cavity, while the positron will be attracted into the bulk. We therefore expect that the mean contact density will be reduced from its value in the hard-wall cavity. We are mainly  interested to see whether these interactions with the wall are sufficient to reduce the mean contact density below its vacuum value.

For the parameters listed above, we obtained  $\langle\eta\rangle=0.0434012$~a.u., still well above the vacuum value of $\eta_0=1/8\pi\approx0.0398$~a.u. Comparing this with the value for the hard-wall cavity, $\langle\eta\rangle=0.0428426$~a.u., we note that $\langle\eta\rangle$ has actually increased; however, the increase is a mere 1.3\% and is likely due to error in extrapolation, the larger $B$-spline box radius causing slower convergence with respect to $l_\text{max}$ and $n_\text{max}$, rather than any physical effect. The  firm  conclusions that can be drawn are that $\langle\eta\rangle$ is still well above its vacuum value and that the change in $\langle\eta\rangle$ from its value in the hard-wall cavity is very small.
This is in contrast to the results of Marlotti Tanzi \textit{et al.} \cite{Tanzi16}, which for  a positron work function in the range $\phi_p=2$--5~eV (${\approx}0.07$--0.2~a.u.) and a cavity radius of $R=0.5$~nm ($\approx$9~a.u.), predicted the contact density to be 10--30\% smaller than its vacuum value (see Fig.~3 in Ref.~\cite{Tanzi16}).

\begin{figure}
\centering
\includegraphics{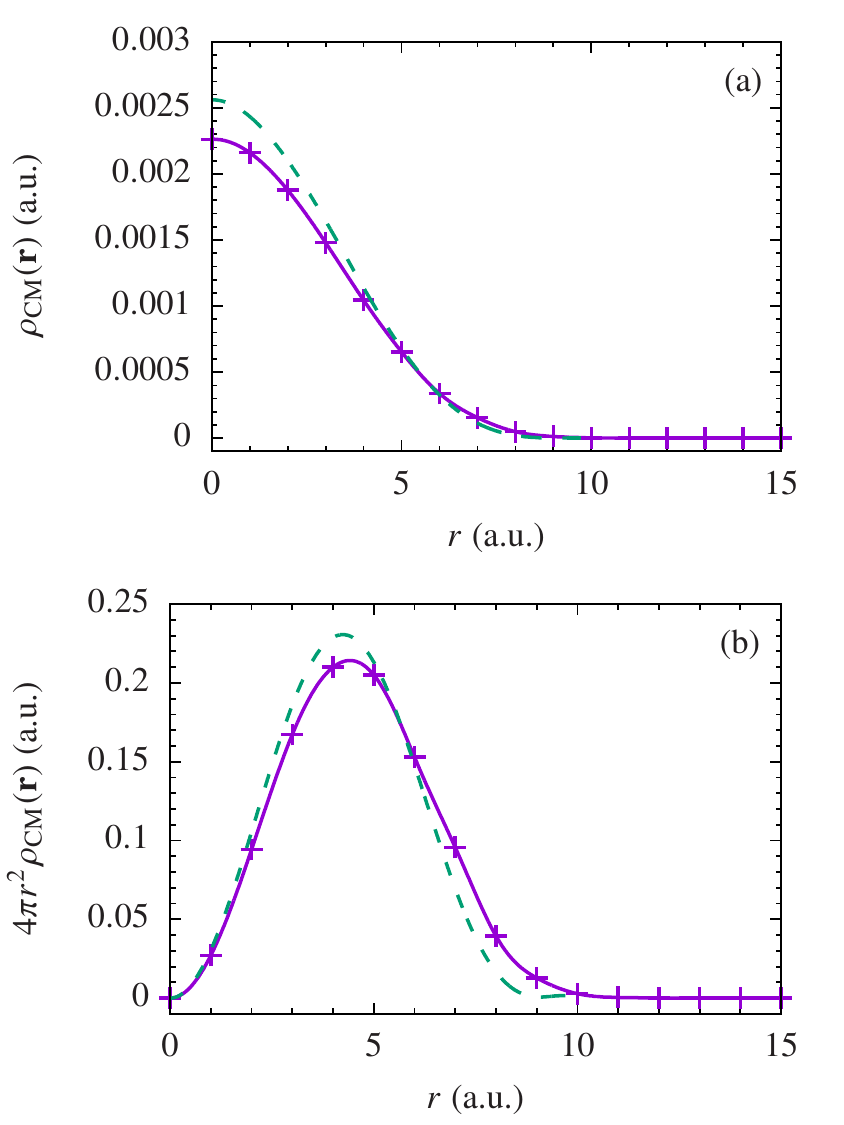}
\caption{\label{fig:COM_density_soft_wall}Center-of-mass density in a cavity of radius $R=10$~a.u. Panel (a) shows $\rho_\text{cm}(\vec{r})$ itself, while panel (b) shows $4\pi r^2\rho_\text{cm}(\vec{r})$. Purple plusses, calculations  with $l_\text{max}=n_\text{max}=20$; solid purple lines, interpolated from calculations  using cubic splines; dashed green lines, interpolated from calculations for hard wall at $R=10$~a.u. (see Fig.~\ref{fig:COM_density_hard_wall}).}
\end{figure}

Figure~\ref{fig:COM_density_soft_wall} shows the center-of-mass density $\rho_\text{cm}(\vec r)$ as a function of $r$ (purple plusses and lines). 
We see that the center of mass is still mostly confined to the region $r<R$, but  there is some penetration into the region $r>R$. 
For comparison, the figure also shows the interpolated center-of-mass density in the hard-wall cavity of radius $R=10$~a.u. (dashed green line, same as solid purple line in Fig.~\ref{fig:COM_density_hard_wall}).
The density of the center of mass at the center of the cavity has decreased by approximately 12\% from the corresponding value in the hard-wall cavity.
This is as expected: the overall ``softening'' of the electron-wall and positron-wall repulsion (in fact, becoming attractive for the positron) allows the Ps to spend more time near the cavity wall.

\begin{figure}
\centering
\includegraphics{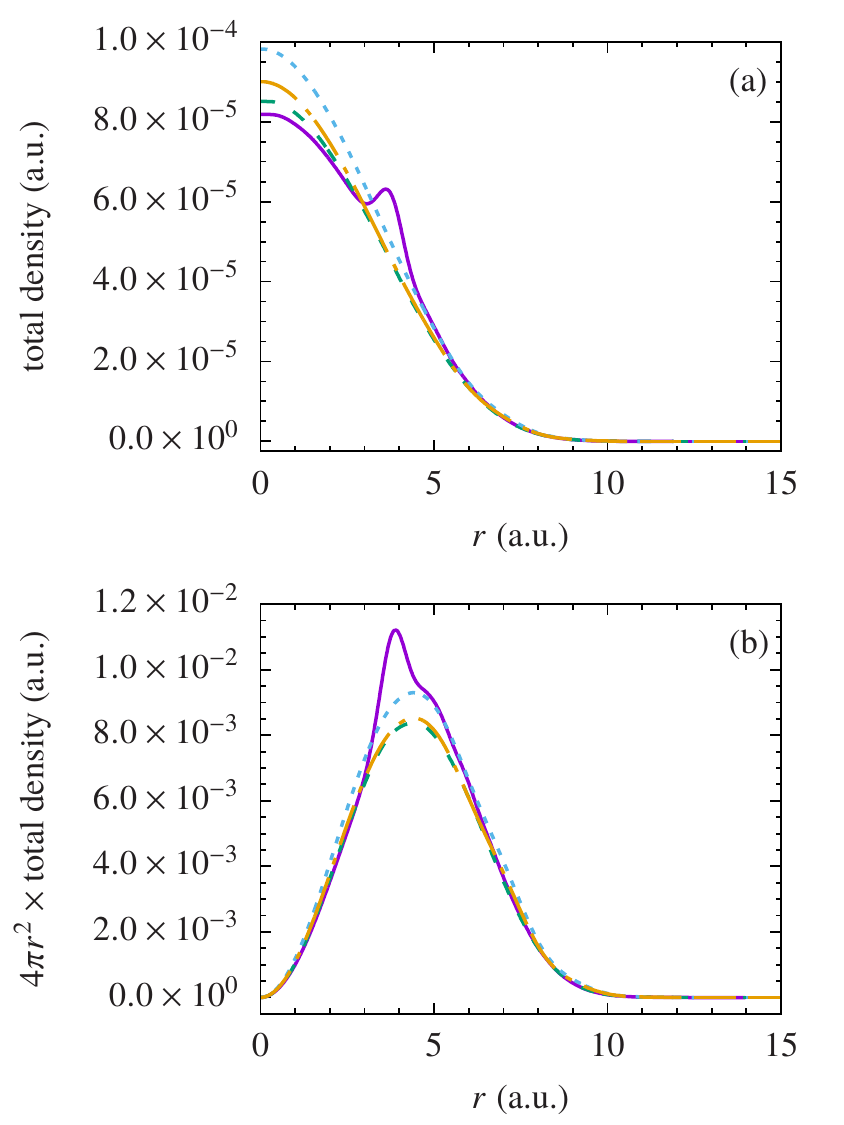}
\caption{\label{fig:total_density_soft_wall}Total density in a  cavity of radius $R=10$~a.u. Panel (a) shows the total density itself, while panel (b) shows $4\pi r^2$ multiplied by the total density. Solid purple curve, direct calculation of $\rho_\text{tot}(\vec{r})$ with extrapolation via Eq.~(\ref{eq:rho_tot_extrp}); dashed green curve, direct calculation of $\rho_\text{tot}(\vec{r})$ with extrapolation via Eq.~(\ref{eq:rho_tot_extrp_alt}); dotted blue curve, $\langle\eta\rangle \rho_\text{cm}(\vec{r})$; dash-dotted orange curve, $\eta_0\rho_\text{cm}(\vec{r})$.}
\end{figure}

Finally, the total density is shown in Fig.~\ref{fig:total_density_soft_wall}.
Once again, extrapolating via Eq.~(\ref{eq:rho_tot_extrp}) produces a bump in $\rho_\text{tot}(\vec{r})$, though the range of values of $r$ it encompasses has moved from $r\approx 2$--3~a.u. for the hard-wall cavity to $r\approx 3$--5~a.u.; it appears that the position of the bump has moved in proportion to the $B$-spline box radius (recall that $R_B=10$~a.u. for the hard-wall cavity, while $R_B=15$~a.u. for the soft-wall cavity). 
 At the center of the cavity,  $\rho_\text{tot}$ is, depending on the type of extrapolation used, approximately 13--16\% less than $\langle\eta\rangle\rho_\text{cm}$, and it is 6--9\% less than $\eta_0\rho_\text{cm}$. This indicates that when the Ps center of mass is at the center of the cavity, the electron-positron contact density is less than both its mean value in the cavity and its vacuum value. It is  plausible that the contact density when the center of mass is at the center of the cavity is less than the mean value throughout the cavity.
Again,  however, it is unlikely that it could truly be smaller than its vacuum value. The observed 6--9\% deficit is probably due to error in the extrapolation, caused in part by the large $B$-spline box radius.
We see that for $r\gtrsim5$~a.u., $\rho_\text{tot}(\vec{r})$, extrapolated using Eq.~(\ref{eq:rho_tot_extrp}) (solid purple line),  is very close to $\langle\eta\rangle\rho_\text{cm}(\vec r)$ (dotted blue curve). This suggests that $\eta\approx\langle\eta\rangle$ in this region. Physically this means that when the Ps approaches the wall, it is gently pushed back into the cavity without the relative motion of the electron and positron being strongly affected.

The above  results call into question the conclusion made by Marlotti Tanzi \textit{et al.} \cite{Tanzi16}.
In their model, the electron is strictly confined to the region $r<R$, while the positron-wall potential is
\begin{equation}
V_p(\mathbf r)=
\begin{cases}
0 &\text{if}\quad r<R+\Delta R , \\
-\phi_p &\text{if}\quad r\ge R+\Delta R,
\end{cases}
\end{equation}
where $\Delta R=0.17$~nm ($\approx$3~a.u.).
This strict confinement of the electron, while allowing the positron to move into the bulk, increases the mean distance between the electron and positron, ultimately reducing the expected contact density below the vacuum value, appearing to justify the experimental data that showing Ps is stretched in most molecular solids.
The ``softer'' confinement of the electron in our model (by means of a Woods-Saxon potential with a negative electron work function) is less drastic and provides a more physical description of the electron-wall interaction in a physical pore, since it uses  a realistic electron work function. The electron is able to penetrate the bulk and remain closer to the positron, and the contact density remains larger than its vacuum value.
Thus, experimental data showing that Ps is stretched in pores of most molecular solids \cite{Consolati88,Consolati88a,Consolati90,Consolati91a,Consolati93,Nagashima01}  require further theoretical investigation.

\section{\label{sec:conclusions}Conclusions}

The effects of confinement on a ground-state Ps atom have
been investigated in three levels of approximation. 

In the first model, we  artificially restricted the maximum separation between the electron and positron.
This had already been investigated \cite{Consolati14}, but we showed that the shift of the electron-positron contact density from its value when the separation is unrestricted diminishes exponentially as the maximum separation becomes asymptotically large. 

In the second model, the Ps moves in a hard-wall spherical cavity. We computed the mean contact density, center-of-mass density, and total density, the latter two of which depend on the position of the Ps center of mass in the cavity. 
We found that the mean contact density is larger than the vacuum value. 
We showed that the contact density is smaller (greater) than its mean value when the center of mass is near the center (wall) of the cavity, which was explained as the Ps being compressed when it collides with the wall.
 
 In the third model, we introduced model electron- and positron-wall potentials to describe Ps confined in a mesoporous material. We found that for a cavity of radius $10$~a.u., an electron work function of $\phi_-=-0.5$~a.u.,  and a positron work function of $\phi_+=0.1$~a.u. (close to reported experimental values for silica \cite{Griscom77,Trukhin92,Nagashima98}), the mean contact density remained above the vacuum value and hardly changed from its value in the hard-wall approximation. This  is in stark contrast to the model of Marlotti Tanzi \textit{et al.} \cite{Tanzi16}, which for a similar cavity radius and positron wave function predicted  the mean contact density to be 10--30\% smaller than its vacuum value. The large discrepancy between our calculation and that of Marlotti Tanzi \textit{et al.} \cite{Tanzi16} arises because we have modeled the electron-wall interaction via an electron work function, while they enforced strict confinement of the electron within the cavity, and they allowed the positron to penetrate freely into the cavity wall by ${\approx}$3~a.u. before changing the potential abruptly to $-\phi_+$. We believe that our model, where the electron and positron move in a cavity with the same effective radius and interact with the wall via realistic work functions, provides a better description of Ps  confined in a pore of a mesoporous material. 
  The conclusion drawn by Marlotti Tanzi \textit{et al.}---that modeling  Ps in a pore by strict confinement of the electron justifies the fact that the contact density is usually measured to be well below the vacuum value \cite{Tanzi16}---is therefore called into question. 
 Although physical pores are not necessarily spherical, our results are still expected to be qualitatively correct.
 Unfortunately, in our approach the contact density and total density converge slowly with respect to the number of  electron and positron basis states included in the Ps wave function. Although we have extrapolated the results, this introduces some uncertainty, most clearly seen by the unphysical ``bumps'' that appear in the total density. Convergence could be aided by using larger numbers of basis states (which requires more computational resources) or by implementing explicitly correlated basis functions. This will reduce the error in the results arising in the extrapolation, but the qualitative conclusions drawn are unlikely to change.

Since our third model predicts that the mean contact density remains above the vacuum value for a physical cavity, 
it fails to explain the results of experiments that found a significant lowering of the contact density from the vacuum value for many materials (see, e.g., Refs. \cite{Consolati88,Consolati88a,Consolati90,Consolati91a,Consolati93,Nagashima01}). The reasons for the discrepancy between theory and experiment are unclear and  warrant further investigation.

\begin{acknowledgments}
We are grateful to D. B. Cassidy for bringing the work of Marlotti Tanzi \textit{et al.} \cite{Tanzi16} to our attention. A.R.S. was supported by the Department for Employment and Learning, Northern Ireland. D.G.G. is funded by ERC grant 804383.
\end{acknowledgments}

\appendix

\section{\label{sec:asymp_shifts}Asymptotic expressions for the energy and relative contact density shifts}

We wish to obtain an asymptotic formula for the energy shift $\Delta E$, for Ps with a maximum electron-positron separation $R$, as $R\to\infty$.
The energy of the Ps is given by Eq.~(\ref{eq:energy_def}), where $n\to1$ as $R\to\infty$. We  put
\begin{equation}
n=1+\epsilon,
\end{equation}
where $\epsilon\to0$ as $R\to\infty$. Using the standard series expansion of the Kummer function \cite{AbramowitzStegun} we find
\begin{equation}
M\left(-\epsilon,2,\frac{R}{1+\epsilon}\right) = 1-\epsilon \sum_{s=1}^{\infty} \frac{R^s}{s(s+1)!}
%M\left(-\epsilon,2,\frac{R}{1+\epsilon}\right) &= 1 - \frac{\epsilon}{2\cdot 1!}R - \frac{\epsilon}{2(2+1)2!}R^2 \nonumber\\
%&\quad {} - \frac{2\epsilon}{2(2+1)(2+2)3!}R^3 + \cdots
\end{equation}
to first order in $\epsilon$. From this we realize that
\begin{equation}
M\left(-\epsilon,2,\frac{R}{1+\epsilon}\right) = 1-\epsilon I,
\end{equation}
where 
\begin{equation}
I\equiv \int_0^{R} \frac{e^t-1-t}{t^2} \, dt.
\end{equation}
We use integration by parts to obtain
\begin{align}
I&=-\frac{e^R}{R} + \frac{1}{R} + 1 + \int_0^R \frac{e^t-1}{t} \, dt \nonumber\\
&= -\frac{e^{R}}{R} + \frac{1}{R} + 1 + \operatorname{Ei}R - \gamma - \ln R,
\end{align}
where 
\begin{equation}
\operatorname{Ei}R = -\int_{-R}^\infty \frac{e^{-t}}{t} \, dt
\end{equation}
is the exponential integral and $\gamma=0.5772\dots$ is the Euler-Mascheroni constant. For large $R$ we may invoke the asymptotic expansion for $\operatorname{Ei}R$ \cite{AbramowitzStegun}:
\begin{equation}\label{eq:Ei_asymp}
\operatorname{Ei}R \approx \frac{e^{R}}{R} \sum_{s=0}^{N-1} \frac{s!}{R^s}.
\end{equation}
Note that this asymptotic series diverges as $N\to\infty$; consequently we retain only the first few terms. The leading terms for $I$ are then
\begin{equation}
I\simeq \frac{e^{R}}{R^2} + \frac{2e^{R}}{R^3} + \frac{6e^{R}}{R^4} + \frac{24e^{R}}{R^5}.
\end{equation}
The value of $\epsilon$ is found from the equation $1-\epsilon I=0$, which yields
\begin{equation}
\epsilon \simeq \left( R^2 - 2R - 2 - \frac{8}{R} \right) e^{-R},
\end{equation}
and thus, using $E=-1/[4(1+\epsilon)^2]$, we obtain Eq.~(\ref{eq:DeltaE_expr}).

We now turn to the contact density. From Eq.~(\ref{eq:1overA2}) we have
\begin{equation}
\frac{1}{\lvert A\rvert^2} = \int_0^R dr \, r^2 e^{-r/(1+\epsilon)} \left( 1-\epsilon\int_0^r dt \, \frac{e^t-1-t}{t^2} \right)^2.
\end{equation}
Remembering that $\epsilon=O(R^2 e^{-R})$,
\begin{equation*}
\int_0^r dt \, \frac{e^t-1-t}{t^2} = O\left(r^{-2} e^r\right)
\end{equation*}
for large $r$, and neglecting terms of order less than $e^{-R}$, this gives
\begin{equation}\label{eq:1overA2_sum}
\frac{1}{\lvert A\rvert^2} = I_1 + I_2 + I_3,
\end{equation}
where
\begin{subequations}
\begin{align}
I_1 &= \int_0^R dr \, r^2 e^{-r}(1+\epsilon r) ,\\
I_2 &= -2\epsilon\int_0^R dr \, r^2 e^{-r} \int_0^r dt \, \frac{e^t-1-t}{t^2} ,\\
I_3 &= \epsilon^2\int_0^R dr \, r^2 e^{-r} \left( \int_0^r dt \, \frac{e^t-1-t}{t^2} \right)^2.\label{eq:I3}
\end{align}
\end{subequations}
Let us consider each of these contributions separately. For $I_1$ we can integrate by parts to obtain
\begin{equation}\label{eq:I1_ans}
I_1 \simeq 2 + \left( 5R^2 - 14R - 14 \right) e^{-R}.
\end{equation}
For $I_2$, changing the order of integration gives
\begin{align}
I_2 &= -2\epsilon\int_0^R dt \, \frac{e^t-1-t}{t^2} \int_t^R dr\, r^2 e^{-r} \nonumber\\
&= 2\epsilon \left[\left( R^2 + 2R + 2\right) e^{-R} I - 2X - 2Y - Z\right],
\end{align}
where
\begin{subequations}
\begin{align}
X &\equiv \int_0^R \frac{1-e^{-t}-te^{-t}}{t^2} \, dt = \frac{1}{R} \left( e^{R} + R - 1 \right),\\
Y &\equiv \int_0^R \frac{1-e^{-t}-te^{-t}}{t} \, dt \nonumber\\
&= -1+\gamma+e^{-R}+E_1(R)+\ln R,\\
Z &\equiv \int_0^R \left( 1-e^{-t}-te^{-t}\right) \, dt = R-2 + (R+2)e^{-R},
\end{align}
\end{subequations}
and
\begin{equation}
E_1(R) \equiv \int_R^\infty \frac{e^{-t}}{t} \, dt.
\end{equation}
Using the asymptotic series \cite{AbramowitzStegun}
\begin{equation}
E_1(R) \approx \frac{e^{-R}}{R} \left( 1- \frac{1!}{R} + \frac{2!}{R^2} - \frac{3!}{R^3} + \cdots \right)
\end{equation}
and neglecting terms of order less than $R^0 e^{-R}$, we obtain
\begin{align}\label{eq:I2_ans}
 I_2 &\simeq 2\big[ {-}R^3 + (5-2\gamma-2\ln R)R^2 \nonumber\\
&\quad{}+ 2(1+2\gamma+2\ln R)(R+1)  \big].
\end{align}
Finally, we come to $I_3$. Using the asymptotic series for the squared integral in Eq.~(\ref{eq:I3}) at large $r$, we have
\begin{align}
I_3 &\sim \epsilon^2 \int_0^R r^2 e^{-r} \left( \frac{e^r}{r^2} + \frac{2e^r}{r^3} + \frac{6e^r}{r^4} \right)^2 \, dr \nonumber\\
&\sim \epsilon^2 \int_0^R r^2 e^{r} \left( \frac{1}{r^4} + \frac{4}{r^5} + \frac{16}{r^6}\right) \, dr.
\end{align}
Making the substitution $r=R-\xi$ and replacing the upper integration limit $R$ by $\infty$, this yields
\begin{align}
I_3 &\sim \epsilon^2 e^R \int_0^\infty \left( R^2-2R\xi + \xi^2\right) e^{-\xi} \Bigg[ \frac{1}{R^4}\left( 1 + \frac{4\xi}{R}+\frac{10\xi^2}{R^2}\right) \nonumber\\
&\quad{}+ \frac{4}{R^5}\left( 1 + \frac{5\xi}{R}\right) + \frac{16}{R^6} \Bigg] \, d\xi \nonumber\\
&\sim\epsilon^2 e^R \left( \frac{1}{R^2} + \frac{6}{R^3} + \frac{34}{R^4}\right).
\end{align}
Neglecting terms of order less than $R^0 e^{-R}$, this yields
\begin{equation}\label{eq:I3_ans}
I_3 \sim \left( R^2 + 2R + 10\right)e^{-R}.
\end{equation}
Combining Eqs.~(\ref{eq:etarel}), (\ref{eq:Deltaetarel}), (\ref{eq:1overA2_sum}), (\ref{eq:I1_ans}), (\ref{eq:I2_ans}), and (\ref{eq:I3_ans}) finally gives Eq.~(\ref{eq:con_dens_approx}).

\section{\label{sec:integrals}Calculation of Coulomb matrix elements and density integrals}

To compute the Coulomb matrix elements (\ref{eq:cou_mat_ele}), we expand the Coulomb potential in Legendre polynomials, viz.,
\begin{equation}
\frac{1}{\lvert \vec{r}_e-\vec{r}_p\rvert} = \sum_{l=0}^\infty \frac{r_<^l}{r_>^{l+1}} P_l (\cos\omega),
\end{equation}
where $r_<=\min(r_e,r_p)$ , $r_>=\max(r_e,r_p)$, $P_l$ is a Legendre polynomial, and $\omega$ is the angle between $\vec{r}_e$ and $\vec{r}_p$, i.e., $\cos\omega=\vec{\hat{r}}_e \cdot \vec{\hat{r}}_p$. Then we separate the radial and angular parts in the single-particle wave functions, integrate over the angular variables, and sum over the magnetic quantum numbers and spins (see, e.g., Ref. \cite{Varshalovich88}), giving
\begin{align}
\langle \nu' \mu' \vert V \vert \mu \nu \rangle =
\sum_l (-1)^{J+l}
\begin{Bmatrix}
 J & l_{\nu'} & l_{\mu'} \\ 
l & l_\mu & l_\nu
\end{Bmatrix} 
\langle \nu' \mu' \Vert V_l \Vert \mu \nu \rangle ,
\end{align}
where $J$ is the angular momentum to which the electron and positron are coupled, and $\langle \nu' \mu' \Vert V_l \Vert \mu \nu \rangle$ is a reduced Coulomb matrix element, defined by
\begin{align}
\langle \nu' \mu' \Vert V_l \Vert \mu \nu \rangle &= 
\sqrt{[l_{\nu'}][l_{\mu'}][l_\mu][l_\nu]}
\begin{pmatrix}
l_\mu & l & l_{\mu'} \\ 0 & 0 & 0
\end{pmatrix}
\begin{pmatrix}
l_\nu & l & l_{\nu'} \\ 0 & 0 & 0
\end{pmatrix} 
\nonumber \\
&\quad{}\times
\int_0^R\!\! \int_0^R P_{\nu'}(r_p) P_{\mu'}(r_e) \frac{r_<^l}{r_>^{l+1}} \nonumber\\
&\quad{}\times P_\mu(r_e) P_\nu(r_p) \, dr_e \, dr_p , \label{eq:red_cou_ele}
\end{align}
where $[l]\equiv2l+1$. The double radial integral in Eq.~(\ref{eq:red_cou_ele}) is evaluated numerically.

To compute the mean contact density (\ref{eq:mean_contact_density}), the $\delta$ function is expanded in Legendre polynomials, viz.,
\begin{equation}\label{eq:B4}
\delta(\vec{r}_e-\vec{r}_p)=\frac{\delta(r_e-r_p)}{r_e^2} \sum_{l=0}^\infty \frac{[l]}{4\pi} P_l (\cos\omega).
\end{equation}
Again, performing the angular integration analytically and summing over the magnetic quantum numbers and spins gives
\begin{align}
\langle\eta\rangle &= \sum_{n_\mu l_\mu n_\nu l_\nu} \sum_{n_{\mu'} l_{\mu'} n_{\nu'} l_{\nu'}} C_{n_\mu l_\mu n_\nu l_\nu} C_{n_{\mu'} l_{\mu'} n_{\nu'} l_{\nu'}} \nonumber\\
&\quad{}\times\sum_l (-1)^{J+l} 
\begin{Bmatrix}
 J & l_{\nu'} & l_{\mu'} \\ 
l & l_\mu & l_\nu
\end{Bmatrix} 
\langle \nu' \mu' \Vert \delta_l \Vert \mu \nu \rangle ,
\end{align}
where
\begin{align}
\langle \nu' \mu' \Vert \delta_l \Vert \mu \nu \rangle &= \frac{[l]}{4\pi} \sqrt{[l_{\nu'}][l_{\mu'}][l_\mu][l_\nu]}
\begin{pmatrix}
l_\mu & l & l_{\mu'} \\ 0 & 0 & 0
\end{pmatrix}
\begin{pmatrix}
l_\nu & l & l_{\nu'} \\ 0 & 0 & 0
\end{pmatrix} 
\nonumber\\
&\quad{}\times 
\int_0^R  P_{\nu'}(r) P_{\mu'}(r)  P_\mu(r) P_\nu(r) \frac{dr}{r^2} .
\end{align}

The center-of-mass density (\ref{eq:cm_dens_int}) is calculated as follows. For $\vec{r}=\vec{0}$, the $\delta$ function can be expanded similarly to Eq.~(\ref{eq:B4}), giving
\begin{align}
\rho_\text{CM}(\vec{0}) &=  \sum_{n_\mu l_\mu n_\nu l_\nu} \sum_{n_{\mu'} l_{\mu'} n_{\nu'} l_{\nu'}} C_{n_\mu l_\mu n_\nu l_\nu} C_{n_{\mu'} l_{\mu'} n_{\nu'} l_{\nu'}} \nonumber\\
&\quad{}\times\sum_l (-1)^{J}
\begin{Bmatrix}
 J & l_{\nu'} & l_{\mu'} \\ 
l & l_\mu & l_\nu
\end{Bmatrix} \nonumber\\
&\quad{}\times \frac{2[l]}{\pi} \sqrt{[l_{\nu'}][l_{\mu'}][l_\mu][l_\nu]}
\begin{pmatrix}
l_\mu & l & l_{\mu'} \\ 0 & 0 & 0
\end{pmatrix}
\begin{pmatrix}
l_\nu & l & l_{\nu'} \\ 0 & 0 & 0
\end{pmatrix} 
\nonumber\\
&\quad{}\times 
\int_0^R P_{\nu'}(r) P_{\mu'}(r)  P_\mu(r) P_\nu(r) \frac{dr}{r^2} .
\end{align}
For $\vec{r} \neq \vec{0}$, the $\delta$ function expands as
\begin{equation}\label{eq:com_delta_expansion}
\delta \left( \frac{\vec{r}_e+\vec{r}_p}{2} - \vec{r} \right) = \frac{\delta(\lvert \vec{r}_e+\vec{r}_p\rvert/2-r)}{r^2} \sum_{l=0}^\infty \frac{[l]}{4\pi} P_l (\cos\omega).
\end{equation}
A difficulty arises in that the $\delta$ function on the RHS of Eq.~(\ref{eq:com_delta_expansion}) also needs to be expanded. For simplicity, we will only consider the case where $J=0$. Then only the $l=0$ term on the RHS of Eq.~(\ref{eq:com_delta_expansion}) is nonzero:
\begin{align}
\delta \left( \frac{\vec{r}_e+\vec{r}_p}{2} - \vec{r} \right) &= \frac{\delta(\lvert \vec{r}_e+\vec{r}_p\rvert/2-r)}{r^2}\frac{1}{4\pi} \nonumber\\
&= \frac{1}{2\pi r^2} \delta(\lvert \vec{r}_e+\vec{r}_p\rvert-2r),\label{eq:expand1}
\end{align}
using the property $\delta(\alpha x)=\delta(x)/\vert\alpha\vert$.
We expand the $\delta$ function on the RHS as
\begin{equation}\label{eq:exp_coeff_gl}
\delta(\lvert \vec{r}_e+\vec{r}_p\rvert-2r) = \sum_{l'=0}^\infty \frac{[l']}{4\pi} g_{l'}(r_e,r_p) P_{l'} (\cos\omega),
\end{equation}
where the expansion coefficients $g_{l'}$ are to be determined. Multiplying both sides of Eq.~(\ref{eq:exp_coeff_gl}) by $P_l(\cos\omega)\sin\omega$, integrating over $\omega$ between 0 and $\pi$, and changing variables to $x\equiv\cos\omega$, we obtain
\begin{equation}
g_l(r_e,r_p) = 2\pi \int_{-1}^1 \delta\left( \sqrt{r_e^2+r_p^2+2 r_e r_p x} -2r \right) P_l(x) \, dx.
\end{equation}
We recall that a general property of the $\delta$ function is
\begin{equation}
\delta[f(x)] = \sum_i \frac{\delta(x-x_i)}{\lvert f'(x_i) \rvert},
\end{equation}
where the $x_i$ are the roots of $f(x)$. In this case we obtain
\begin{equation}
g_l(r_e,r_p) = \frac{4\pi r}{r_e r_p} \int_{-1}^1 \delta\left( x - \frac{4r^2-r_e^2-r_p^2}{2r_er_p}\right) P_l(x) \, dx ,
\end{equation}
which gives
\begin{equation}\label{eq:gl_final}
g_l(r_e,r_p) = 
\frac{4\pi r}{r_e r_p} P_l \left( \frac{4r^2-r_e^2-r_p^2}{2r_er_p}\right) 
\end{equation}
if $\lvert r_e-r_p \rvert < 2r < r_e+r_p$, and $g_l(r_e,r_p)=0$ otherwise. 
Combining Eqs.~(\ref{eq:cm_dens_int}), (\ref{eq:expand1}), (\ref{eq:exp_coeff_gl}), and (\ref{eq:gl_final}), and noting that we require $l_\mu=l_{\mu'}=l_\nu=l_{\nu'}(\equiv l)$ for $J=0$, we obtain
\begin{align}
\rho_\text{CM}(\vec{r}) &=  \sum_{n_\mu n_\nu n_{\mu'}  n_{\nu'} } \sum_l C_{n_\mu l n_\nu l} C_{n_{\mu'} l n_{\nu'} l} \nonumber\\
&\quad{}\times 
\frac{1}{2\pi r}\int_0^R \!\! \int_{\lvert 2r-r_p\rvert}^{\min(2r+r_p,R)} P_{\nu'}(r_p) P_{\mu'}(r_e)\nonumber\\ 
&\quad{}\times P_l\left( \frac{4r^2-r_e^2-r_p^2}{2r_er_p}\right)
 P_\mu(r_e) P_\nu(r_p)  \frac{dr_e}{r_e}\frac{dr_p}{r_p} .
\end{align}

\section{\label{sec:hard_fit}Dependence of mean contact density on cavity radius for a hard-wall cavity}

The type of fit used in Eq.~(\ref{eq:confit}) can be justified by the following simple analysis.
The total Ps wave function in the cavity is
\begin{equation}
\Psi_{0+}(\vec{r}_e,\vec{r}_p) = \frac{A}{R_\text{cm}} \sin \left(KR_\text{cm}\right) Y_{00}\left(\vec{\hat{R}}_\text{cm}\right)\psi_{1s}(\vec{r}_e-\vec{r}_p),
\end{equation}
where $A=\sqrt{2/R}$ is the normalization constant, $K\approx\pi/R$ is the center-of-mass momentum, $\vec{R}_\text{cm}=(\vec{r}_e+\vec{r}_p)/2$ is the position of the center of mass, and $\psi_{1s}$ is the internal wave function for Ps($1s$).
We assume that if the center of mass is near the center of the cavity, the contact density takes its vacuum value $\eta_0$, while if the center of mass is near the wall, the contact density takes a different (higher) value  $\eta_w$. We define the center of mass to be ``near the wall'' if it is within a thin shell of thickness $\xi \ll R$ next to the wall, i.e.,
\begin{equation}
\eta = 
\begin{cases}
\eta_0 & \text{if} \quad 0 \leq R_\text{cm} <R-\xi , \\
\eta_w & \text{if} \quad R-\xi \leq R_\text{cm} < R.
\end{cases}
\end{equation}
The mean contact density throughout the cavity is then
\begin{equation}\label{eq:prob}
\langle\eta\rangle = \eta_0 P_0 + \eta_w P_w = \eta_0 + P_w(\eta_w - \eta_0) ,
\end{equation}
where $P_w$ is the probability that the center of mass is near the wall, and $P_0 $ is the probability that the center of mass is not near the wall, with $P_0+P_w=1$. We can explicitly calculate $P_w$:
\begin{equation}
P_w = A^2 \int_{R-\xi}^R \sin^2 KR_\text{cm} \, dR_\text{cm} = \frac{1}{R} \left( \xi - \frac{R}{2\pi}\sin\frac{2\pi\xi}{R} \right) .
\end{equation}
Using the Maclaurin expansion of the sine function gives
\begin{equation}\label{eq:pw_exp}
P_w = \frac{2\pi^2}{3} \left(\frac{\xi}{R}\right)^3 - \frac{2\pi^4}{15} \left(\frac{\xi}{R}\right)^5 + O\left[ \left(\frac{\xi}{R}\right)^7\right].
\end{equation}
Combining Eqs.~(\ref{eq:prob}) and (\ref{eq:pw_exp}) finally gives
\begin{equation}
\langle\eta\rangle = \eta_0 + \frac{2\pi^2}{3} (\eta_w-\eta_0) \left(\frac{\xi}{R}\right)^3 - \frac{2\pi^4}{15} (\eta_w-\eta_0) \left(\frac{\xi}{R}\right)^5 + \cdots,
\end{equation}
which is the form used in Eq.~(\ref{eq:confit}).

%\bibliography{Ps_contact_density_bib}

\begin{thebibliography}{75}
\bibitem [{MTE()}]{MTE_footnote}%
  \BibitemOpen
  \href@noop {} {}\bibinfo {note} {This modification of the Tao-Eldrup model is
  valid even for small pore radii, but self annihilation has a negligible
  contribution for pore radii ${\lesssim}1$~nm \cite{Ito99}.}\BibitemShut
  {Stop}%
\end{thebibliography}


\end{document}